\documentclass[aps,twocolumn,showpacs,preprintnumbers,prl,superscriptaddress,10pt,nofootinbib]{revtex4-1}

\usepackage{graphicx}
\usepackage{bm}
\usepackage{natbib}
\usepackage{comment}
\usepackage{ulem} 

\newcommand{\approptoinn}[2]{\mathrel{\vcenter{
          \offinterlineskip\halign{\hfil$##$\cr
        #1\propto\cr\noalign{\kern2pt}#1\sim\cr\noalign{\kern-2pt}}}}}

\usepackage{color}

\def\red{\textcolor{red}}
\def\bl{Babcock--Leighton}
\newcommand\mnras{{MNRAS}}%
\newcommand\aap{{Astrophys. J}}%
\newcommand\solphys{{Sol Phys}}%

\newcommand{\Eq}[1]{Equation~(\ref{#1})}

\newcommand{\Fig}[1]{Figure~\ref{#1}}

\newcommand{\mps}{m~s$^{-1}$}

\graphicspath{{./fig/}{./png/}}

\def\bl{Babcock--Leighton}
\def\mc{meridional circulation}
\def\we{Waldmeier effect}

\begin{document}
\title{Toroidal flux loss due to flux emergence explains why solar cycles rise differently but decay in a similar way}
\medskip
\author{Akash Biswas}
\affiliation{Department of Physics, Indian Institute of Technology (Banaras Hindu University), Varanasi 221005, India}
\author{Bidya Binay Karak}
\email{karak.phy@iitbhu.ac.in}
\affiliation{Department of Physics, Indian Institute of Technology (Banaras Hindu University), Varanasi 221005, India}

\author{Robert Cameron}
\affiliation{Max Planck Institute for Solar System Research, Justus-Von-Liebig-Weg 3, D-37077, G\"ottingen, Germany}
\date{\today}

\begin{abstract}


A striking feature of the solar cycle is that at the beginning, sunspots appear around mid-latitudes, and over time the latitudes of emergences migrate towards the equator.
The maximum level of activity (e.g., sunspot number) varies from cycle to cycle.
For strong cycles, the activity begins early and at higher latitudes with wider sunspot distributions than for weak cycles. The activity and the width of sunspot belts increase rapidly and begin to decline when the belts are still at high latitudes. 
Surprisingly, it has been reported that in the late stages of the cycle the level of activity (sunspot number) as well as the  widths and centers of the butterfly wings all have the same statistical properties independent of how strong the cycle was during its rise and maximum phases. 
We have modeled these features using a Babcock--Leighton type dynamo model and show that the flux loss through magnetic buoyancy is an essential nonlinearity in the solar dynamo. 
Our study shows that the nonlinearity is effective if the flux emergence becomes efficient at the mean-field strength of the order of $10^4$~G in the lower part of the convection zone.
\end{abstract}
\maketitle

\begin{figure}
\centering
\includegraphics[scale=0.54]{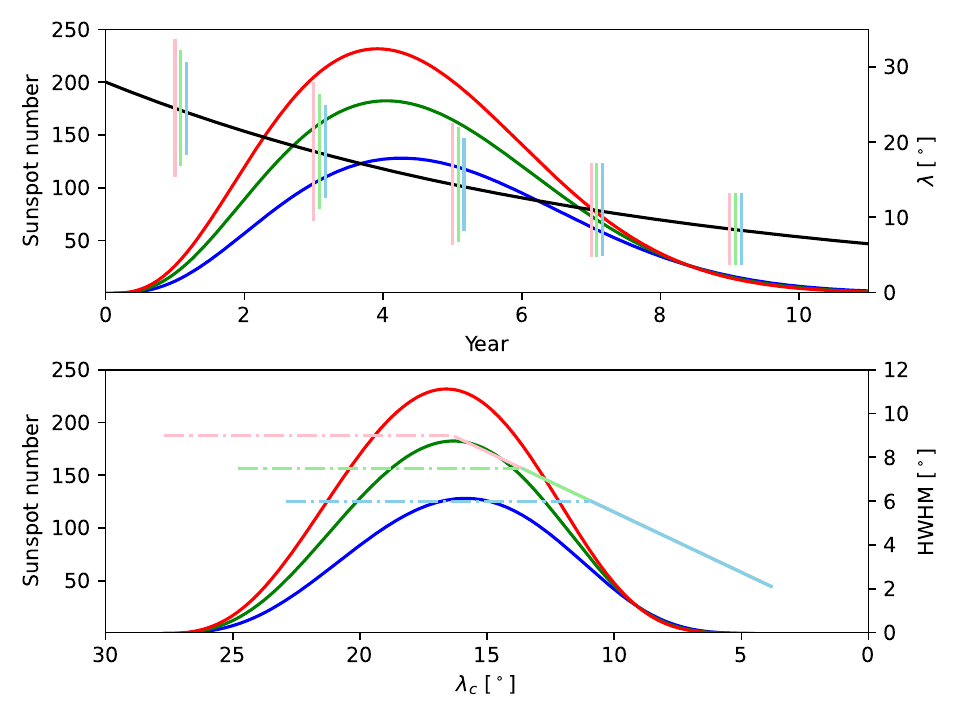}
\caption{Illustration of the behaviour of cycles of different strengths. Upper panel:
The three coloured curves show the sunspot number for cycles of weak (blue), moderate (green) and high (red) levels of activity. The curves are based on the empirical curve-fitting
of Hathaway et al \cite{1994SoPh..151..177H}.
The sunspots appear in butterfly wings, and the central latitude of the butterfly wings drift towards the equator in a way which is independent of cycle strength (black curve)\cite{2011SoPh..273..221H}. Finally, the width of the butterfly wings (light vertical lines) depends on the cycle strength early in the cycle but is independent of the cycle strength late in the cycle  \cite{CS16}. The properties of the butterfly diagram in the late phase of all the cycles are the same, i.e., all cycles die in the same way although they have different properties during the rise phase of the cycle \cite{CS16}.
Lower Panel: because all cycles drift towards the equator in the same way, the central latitude of the butterfly wings can be used as the independent variable rather than time from the start of the cycle.
The solid light lines indicate the width of the butterfly wings as a function of latitude. The dashed light lines are the upper limit for the width during the early phases of the cycle: the width increases as the cycle progresses until it reaches this level. Thereafter the cycle begins to decline.
}
\label{fig:ill}
\end{figure}

Solar activity, as measured for example by the number of sunspots on the solar surface, takes place in 11-year cycles. The strength of each cycle (e.g., the maximum number of sunspots) varies from cycle to cycle. 
Strong and weak cycles have systematic differences: compared to weak cycles, strong cycles begin at higher latitudes, rise more rapidly, reach their maxima earlier and consequently have a longer decline phase \cite{W35}. 
This is known as the Waldmeier effect
which has also been confirmed in the 
reconstructed solar activity data over millennial time scale \citep{Usos21a}.
A stronger constraint is that all cycles behave in the same way towards the end of the cycle \citep[][hereafter CS16]{CS16}. This is illustrated in Figure~\ref{fig:ill}, and we refer the reader to the work by Hathaway \cite{2011SoPh..273..221H} and CS16 for the observational analysis.
The fact that the cycles all have the same properties (amplitude, spatial distribution) in the late phases of the cycle despite having different amplitudes in the early phase of the cycle indicates a nonlinearity acting as the cycle progresses. The nonlinearity is particularly important because during the late phase of the cycle sunspots emerge closer to the equator and hence contribute more to the build-up of Sun's polar fields \citep{2020JSWSC..10...50P}.
 (The magnetic flux of the leading polarity of the bipolar magnetic regions (BMRs) that emerge closer to the equator gets easily carried across the equator by small-scale convective motions. 
 It is the cross-equatorial cancellation of this flux that changes the net flux in each hemisphere and the polar fields at the end of each cycle 
 \citep{2004SoPh..222..345D}).
The nonlinearity hence leads to the change in polar field from one cycle to the next being only weakly dependent of the cycle strength \cite{J20}.  

In recent years, \bl\ dynamo models have 
explained many features of the irregular solar cycle 
\cite{Char20}.
For example, \citet{KC11} explained the Waldmeier effect
 using a \bl\ type flux transport dynamo model.
More recetly, \citet{MKB17} reproduced the basic  correlations described above using a \bl\ dynamo model. However, why they were successful in this regard was not discussed.
Here, we shall employ a \bl\
type dynamo model
to explain the features reported by \citet{W55} and CS16.
We shall further show that in order to reproduce these features, some constraints on the nonlinearity in terms of the equipartition field strength, and the variation of meridional flow 
can be inferred.

For our study, we use an updated version of the code SURYA \citep{NC02, CNC04}, which solves the following equations:

\begin{equation}
\frac{\partial A}{\partial t} + \frac{1}{s}({\bm v_p} \cdot {\bf \nabla})(s A)   = \eta_p\left(\nabla^2 - \frac{1}{s^2}\right)A + \alpha B,
\label{eq:pol}
\end{equation} 

\begin{eqnarray}
\frac{\partial B}{\partial t} + \frac{1}{r}\left[\frac{\partial (rv_rB)}{\partial r}+ \frac{\partial (v_\theta B)}{\partial \theta}  \right] = \eta_t\left(\nabla^2 - \frac{1}{s^2}\right)B 
\nonumber\\
+ s({\bm B_p} \cdot {\bf \nabla})\Omega + \frac{1}{r}\frac{d\eta _t}{dr}\frac{\partial (rB)}{\partial r}
\label{eq:tor}
\end{eqnarray}
where $r$ is radial distance from the center of the Sun, $\theta$ is the co-latitude, 
$B(r,\theta)$ is the toroidal field and $A(r,\theta)$ is the $\phi$ component of the magnetic vector potential of the poloidal magnetic field ${\bm B_p}$, $s=r\sin{\theta}$, 
${\bm v_p}= v_r \hat{\bm{r}}+ v_\theta \hat{\bm{\theta}}$ is the meridional circulation,  
$\Omega (r,\theta)$ is the local rotation rate, $\eta_p (r)$ and $\eta_t (r)$
are the turbulent diffusivities for the poloidal and toroidal fields. The coefficient $\alpha (r,\theta)$ captures the generation of the poloidal field near the solar surface through the decay 
and dispersal of the tilted BMRs (\bl\ process) in our axisymmetric model.

As the profiles of all parameters are given in many publications \citep[e.g.,][]{Kar10,KMB18,CNC04, KC11, KC16}, we 
discuss them in the Supplemental Material \citep{Suppli} 
except 
$\alpha$ and magnetic buoyancy,
which play important role in 
the present work.

In the regular model, the $\alpha$ 
has following form
\begin{eqnarray}
\alpha(r,\theta) = \frac{\alpha_0}{4}\cos{\theta}\left[1+ \mathrm{erf}\left(\frac{r-0.95R_\odot}{0.03R_\odot}\right)\right]
\nonumber\\
 \times \left[1- \mathrm{erf}\left(\frac{r-R_\odot}{0.03R_\odot}\right)\right]
\end{eqnarray}

where $\alpha_0= 30~$\mps. 

In the \bl\ dynamo, generation of poloidal field involves
some randomness, primarily due to  scatter around
Joy's law and the random flux emergences. 
To capture these effects in our model, we include stochastic noise in the above $\alpha$ by replacing $\alpha_0$ by $\alpha_0 \left[ 1 + \sigma(t, \tau_{\rm corr})\right]$,
where $\sigma(t, \tau_{\rm corr})$ is a random number drawn from a uniform distribution in the range $[-1,1]$ and $\tau_{\rm corr}$ is the coherence time over which the value of $\alpha$ is constant. In all simulations,
$\tau_{\rm corr} = 1$ month (typical lifetime of sunspots).
With these parameters, the variations in the solar cycle remain 
consistent with what is seen in the last 300 years.



We take the profile for \mc\ to be of the same functional form as in \citep{CNC04} with one change in the parameters: we take $\epsilon = 3$ instead of $\epsilon = 2$ as given there. 
This change makes the meridional circulation to increase slightly at the lower latitudes and decrease at higher latitudes 
compared to the previous profile \citep{CNC04}; see \citep{Suppli} for details. 
With this slight modification in the flow, the equatorward migration of sunspot belt is in better agreement with the observations.

Finally, we describe the only nonlinearity that has been included in the model.
The nonlinearity concerns
the time at which flux is assumed to emerge. 
Every 0.6 days, the code checks the amplitude of the toroidal field 
at each grid point above the base of the convection zone (BCZ). If the value is above a to-be-described
critical value ($B_c$) 
then it is assumed that half of the flux emerges --
its value at that grid point is halved and this half is added to the toroidal field near the surface at the same latitude. This is how the sunspot eruptions are modelled and this mechanism ensures that with each eruption a part of the toroidal flux from the CZ is lost. 
When the threshold is met, the toroidal flux is removed from the lower CZ so that it is no longer available to emerge at lower latitudes. This flux-loss associated with a threshold is the only nonlinearity we have included.
Such a nonlinearity tends to make all cycles behave in the same way during the late phases of the cycles because in the late phase a lot of the toroidal flux associated with a strong cycle has already been removed. 
It acts to saturate the dynamo because emergences at high-latitudes tend to be ineffective in terms of changing the polar fields \cite{2020JSWSC..10...50P}.






\begin{figure}
\centering
\includegraphics[scale=0.1]{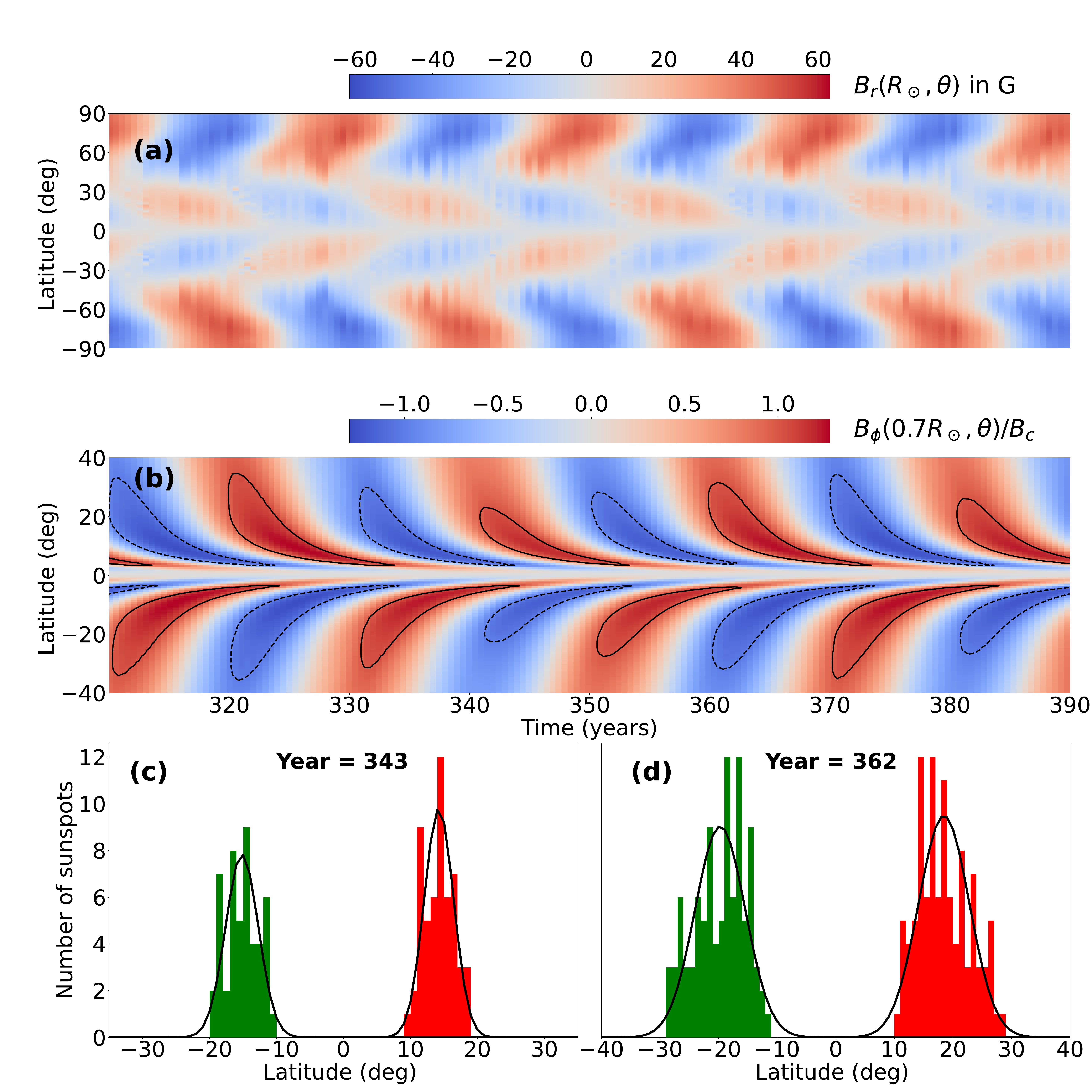}
\caption{A typical portion from our simulation. 
Latitudinal distributions of (a) the surface radial field and 
(b) the toroidal field at BCZ (the contour represents $B=B_c$).
(c) and (d) 
show the annual latitudinal distributions of sunspots for the years: 343 and 362, 
respectively and the fitted Gaussian profiles (black curves).}
\label{fig:bfly}
\end{figure}

\begin{figure}
\centering
\includegraphics[scale=0.35]{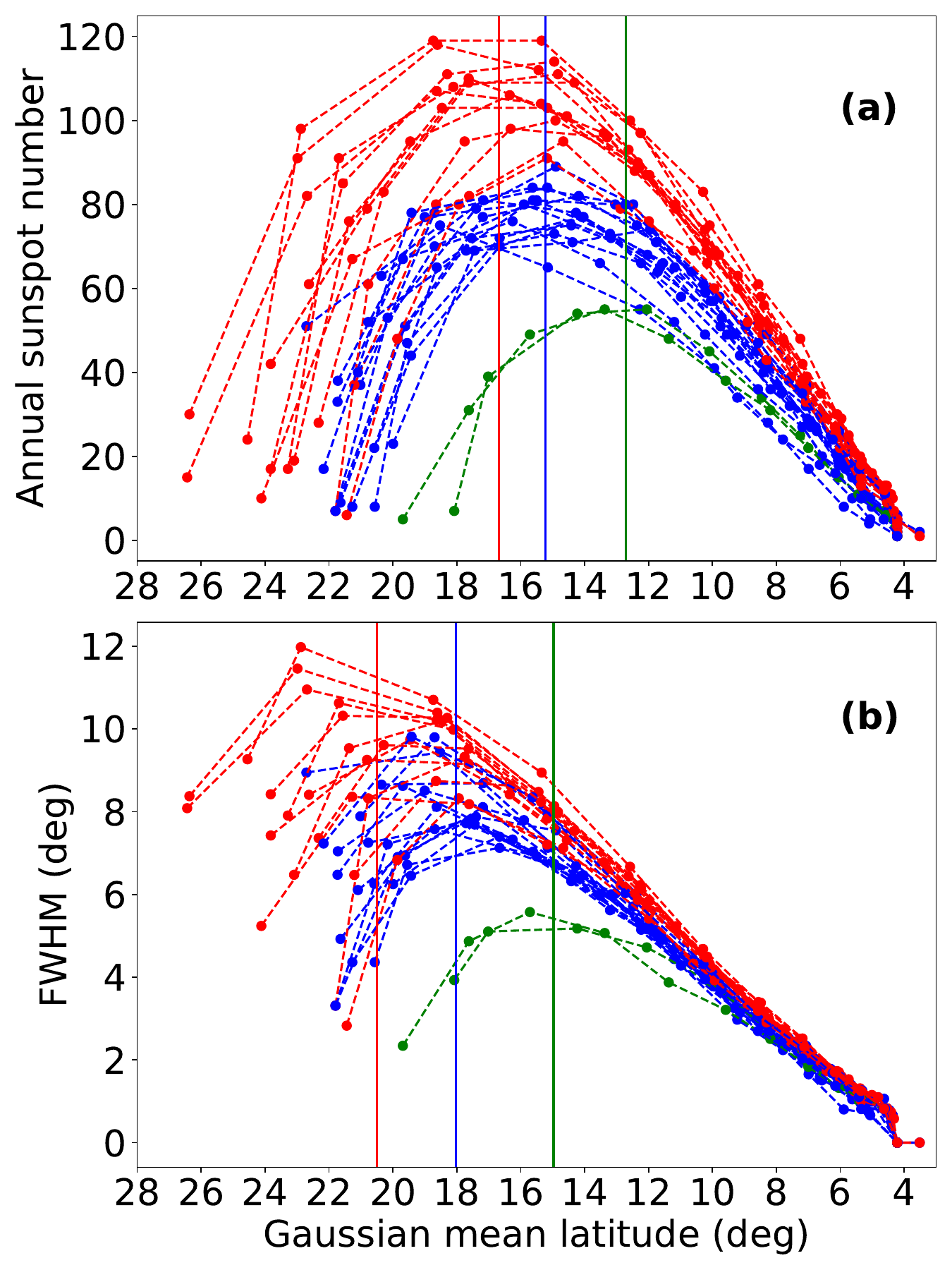}
\caption{(a) 
Model (pseudo) sunspot number for each year (based on the number of flux emergences) 
as a function of the central latitude of the Gaussian distribution. Different curves correspond to different cycles. (b) Same as (a) but for the distribution width (FWHM). Cycles begin at the left side of the plots and migrate to the right as they evolve.
Vertical lines guide the average of the Gaussian-mean latitudes for strong (red), moderate (blue) and weak (green) cycles at their peaks. 
}
\label{fig:CS}
\end{figure}

We now discuss the results of the dynamo simulation.
The latitudinal distribution of the surface radial field for eight cycles from a long run of 40 cycles is shown in \Fig{fig:bfly}(a). We observe that this plot resembles the basic features of the solar cycle reasonably well \citep{Hat15,Char20}. 
We find that when $B_c = 0.8 \times 10^4$~G, the strength of the
radial field on the surface is in agreement with the observed range \cite{Mord22} and the subsurface flux loss through magnetic buoyancy becomes consistent with the observed flux loss through BMR emergences \citep{CS20}. 
We use this value of $B_c$ for the rest of this Letter.

{
We compute the theoretical sunspot number by tracking the latitudes of eruptions of the toroidal flux (where the bottom field surpasses $B_c$) and with this number, we  perform the same analyses as done in CS16 for the observed data.
}
In \Fig{fig:bfly}(c-d), we show that the latitudinal distribution of annual 
(model) 
sunspot emergence can be approximated with a Gaussian distribution. The solar cycle variations of the total annual sunspot number and the width of sunspot distribution with the central latitudes of distribution are shown in \Fig{fig:CS}. These behaviors largely agree with the observations (CS16).

We now explain these results based on the dynamo theory.
In our model, cycles vary in strength due to the fluctuations introduced in the
\bl\ $\alpha$ term.  Suppose for some time $\alpha$ is large in a cycle, then more poloidal field will be produced in that cycle. 
The poloidal field is sheared by the differential rotation to produce the next cycle's toroidal field (we note that the shear is strongest at about $55^\circ$ \citep{How09}). 
In our model, this toroidal field emerges once a critical threshold is reached, and gives rise to the (pseudo)
sunspots on the solar surface.  
Hence, a high value of $\alpha$ in a cycle produces a strong poloidal field. 
This generates a strong toroidal field for the next cycle and 
the eruption condition ($B > B_c$) is satisfied at earlier times when the field is at high latitudes. 
This explains why, for a strong cycle,
sunspot starts appearing at high latitudes and the width of the latitude belt is large. 
On the other hand, when a cycle is weak, the model takes a long time for the toroidal field to satisfy the spot-eruption condition 
by that time the \mc\ drags the toroidal field towards the low latitudes.
Hence, in a weak cycle, the sunspot belt begins at lower latitudes. The band of the sunspot latitudes is also narrow when the toroidal field is weak 
 (as the spot eruption condition is satisfied only in a narrow latitude band). 


Now we consider the effect of the loss of toroidal flux due to flux emergence \citep{CS20}.
Again we consider a strong cycle for which emergence begins early in the cycle at high latitudes. 
Each emergence reduces the subsurface toroidal flux \citep{CS20} so that a strong cycle which has many early flux emergences rapidly loses toroidal flux until 
the subsurface mean field strength is just above $B_c$. Thereafter 
the cycle begins to decline and the toroidal flux is advected equatorward by the meridional flow while maintaining a strength just above $B_c$. 
The decrease in the meridional flow as we approach the equator 
will tend to increase the subsurface field strength, but this will be compensated for by further flux loss due to flux emergence. 

On the other hand, in a weak cycle, there will be very few eruptions early in the cycle, and the field strength will simply build up as the flux is advected towards the stagnation point at the equator. At some point however the mean field will be comparable to $B_c$ and the cycle will enter its decline phase. During this phase, the situation is entirely the same as for a strong cycle: the increase in the field strength as the flux builds up near the stagnation point is balanced by the loss of flux due to emergence,
and the mean field strength is kept around $B_c$. 
A nonlinear process which causes flux emergence rates to be enhanced when the mean field exceeds $B_c$ can 
explain why all cycles decline in the same way as evident from the right part of the curves in \Fig{fig:CS}.
The value of $B_c=0.8 \times 10^4$~G was chosen so that the model matches  the observed range of the radial magnetic field on the solar surface and the amount of flux loss.
Incidentally, this value is close to the 
equipartition field strength at BCZ ($\sim 10^4$~G). Although we did not choose the value of $B_c$ keeping this in mind, 
it is interesting to note that 
just by constraining our model parameters through observations, we get a value of $B_c$, which is close to the equipartition field strength.  
A nonlinearity is expected because at this field strength
 the magnetic field has an energy density comparable to the kinetic energy density of the turbulent motions.


One quantitative discrepancy between our model presented in \Fig{fig:CS}(b) with respect to the observations (CS16) is that in the latter,
we observe that as soon as the distance between the center of the Gaussian and the equator is roughly equal to the FWHM, all cycles begin to decay. 
The width of the butterfly wings is substantially smaller in our model; see \Fig{fig:CS}(b). 
In this Letter, we have been interested in investigating the role of flux loss in combination with a threshold for the emergence of active regions.
For this pupose we have used a simple sharp threshold which involves two parameters $B_c$ and the fraction of flux which emerges when an emergence takes place. 
A better match with the observed widths might be possible if the flux emergence recipe had more degrees of freedom, however the basic physical idea is captured with this simple threshold.
An additional possibility is  that the butterfly wing widths are broadened by convective buffeting of the rising flux tubes before they reach the surface.

\begin{figure}
\centering
\includegraphics[scale=0.38]{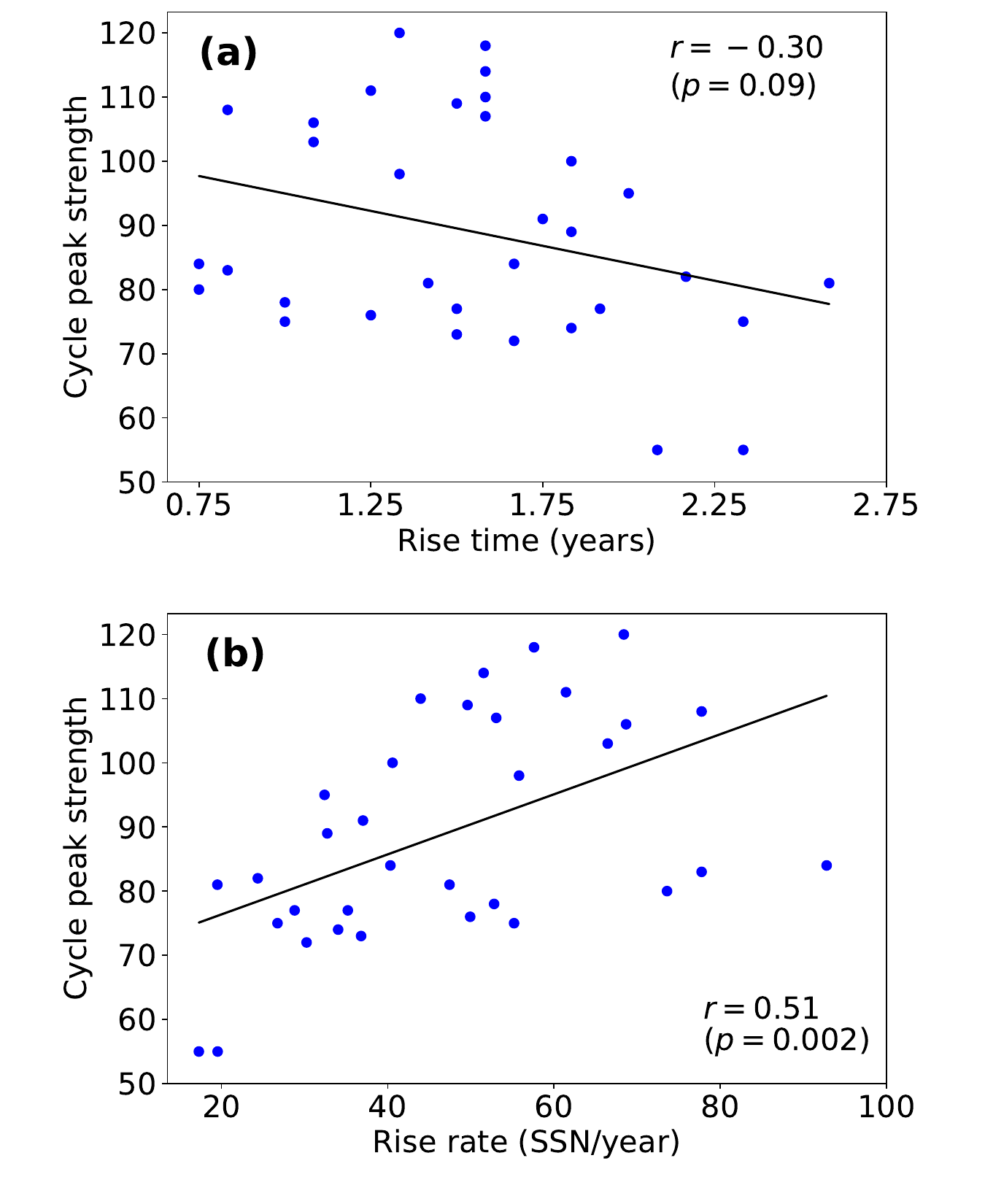}
\caption{
The scatter plots between the peak of sunspot cycle (amplitude) and (a) the rise time and (b) the rise rate, i.e., WE1 and WE2, respectively.
}
\label{fig:WE}
\end{figure}


We have also checked 
the \we\
i.e., WE1 and WE2 \citep{Suppli}. We find linear (Pearson) correlation coefficients $-0.30$ and $0.51$ for these two cases (\Fig{fig:WE}).  Thus the classical \we\ is also reproduced in this simulation. 
Earlier \citet{KC11} showed that fluctuations in \mc\ are needed to reproduce WE1. 
Our result differs from this because WE1 is a weak anti-correlation and is sensitive to the way in which the data (and simulations) are treated.
We find that a weak anti-correlation in WE1 exists even without including fluctuations in the \mc. When fluctuation in \mc\ is introduced, it enhances this anti-correlation. However, it leads to the model disagreeing with the fact that 
the widths of the sunspot latitude bands of all cycles are the same function of  the central latitudes of the sunspot bands (CS16; \Fig{fig:CS}).
Hence, observational features of sunspot cycle \citep{W55} as analyzed in CS16 suggest
that there was no large variation in the deep \mc\ in the past 300 years.

 In this Letter we have focused on explaining why all cycles are statistically the same in there decay phase. The 
dynamo simulation can be modified to include other effects such as a time delay between the start of the rise of a flux tube and when it emerges at the surface. This delay time can be of the order of months
\citep{FFM94}. Such a delay can affect the properties of the dynamo cycles \citep{Jouve10, Fournier18}. We show
\citep{Suppli} that the inclusion of such a delay in our model does not substantially affect the conclusions of this Letter, although it does have a weak effect on the latitudes at which the emergences take place.


In conclusion, we have shown that the main features of the latitudinal distribution of sunspots as reported in \citet{W55} and CS16 are reproduced in a \bl\ type flux transport dynamo model with stochastic fluctuations in the poloidal field source.
We find that a constant equatorward flow near BCZ and a reduction of toroidal field due to flux emergence are essential to reproduce these results.
Further, 
the critical strength of the mean magnetic field for the flux emergence through buoyancy is found to be of the order of $10^4$ G. This is about the equipartition value where the magnetic energy density is equal to the kinetic energy density of the turbulent convective motions. 
\\\\

\begin{center}
\large\textbf{Supplemental Material}
\end{center}

\subsection{S1. Details of the Model Parameters}

The profiles of all the ingredients of our dynamo model have been discussed in many previous publications \citep[e.g.,][]{CNC04, Kar10, KC11, KC16}. However, we discuss the profile of \mc\  as it has been slightly modified here. 

The meridional circulation is obtained from a stream function $\psi$,
such that $\rho {\bm v_p} = \nabla \times [\psi (r, \theta) \hat{\mbox{\boldmath $\phi$}}],$
where
$\rho = C \left( \frac{R}{r} - 0.95 \right)^{3/2}, \label{rho}$
and
\begin{eqnarray}
    \psi r \sin{\theta} = \psi_0(r-R_p)\sin{\left[\frac{\pi(r-R_p)}{(R_\odot - R_p)}\right]}\{1-e^{-\beta_1\theta^\epsilon}\}
    \nonumber\\
    \{1-e^{-\beta_2(\theta-\pi/2)}\}e^{((r-r_0)/\Gamma)^2}
\end{eqnarray}
Here the values of the parameters are:
$C = 0.407$~kg~m$^{-3}$, $\beta_1 = 1.5 \times 10^{-8}$~ m$^{-1}$, $\beta_2 = 1.3 \times 10^{-8}$~m$^{-1}$, $r_0 = (R_\odot - R_b)/3.5$, $\Gamma = 3.1 \times 10^8$~m, $R_p = 0.635R_\odot$, $R_b = 0.55R_\odot$.

\begin{figure*}


\includegraphics[scale=0.3]{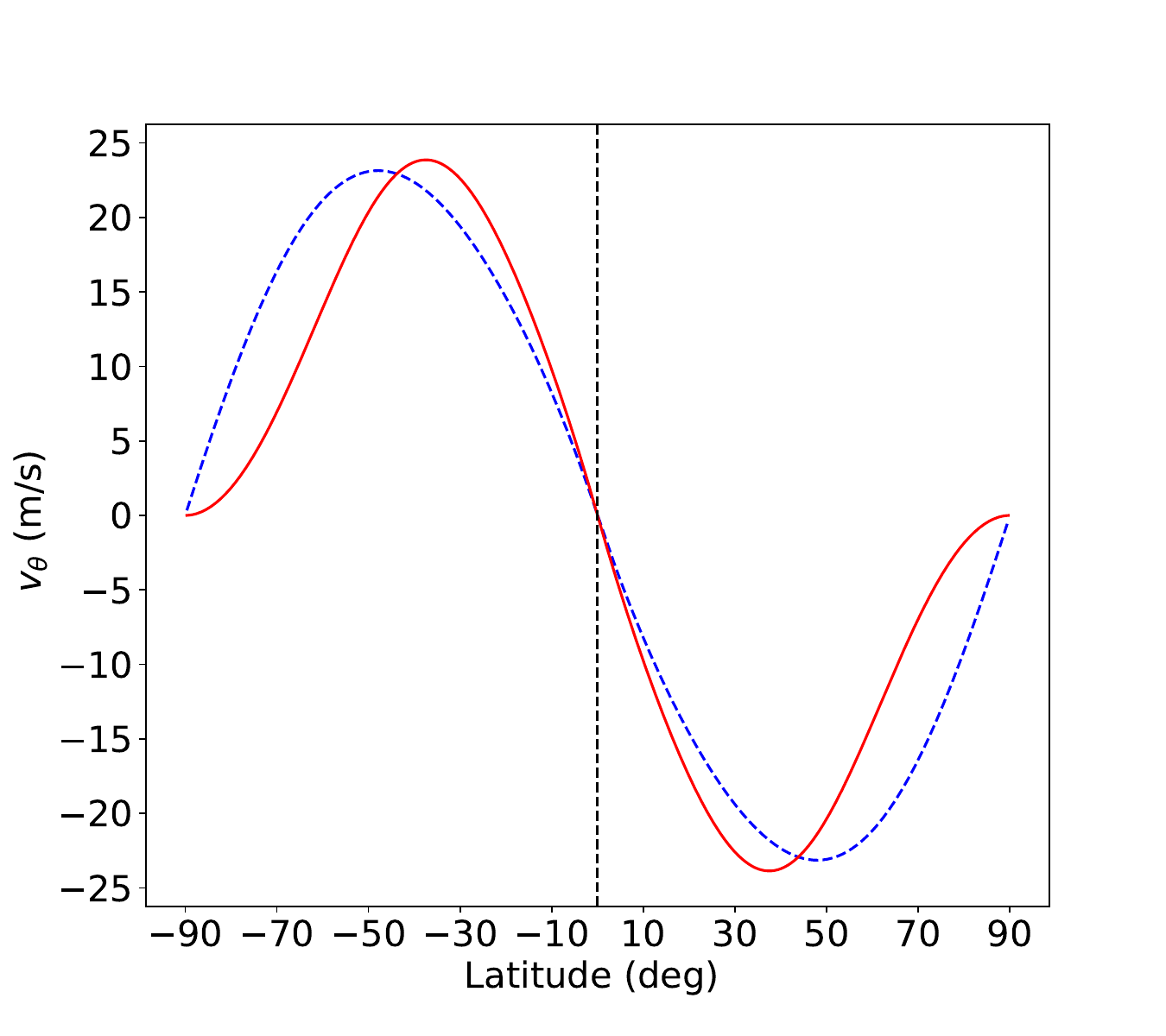}
\includegraphics[scale=0.19]{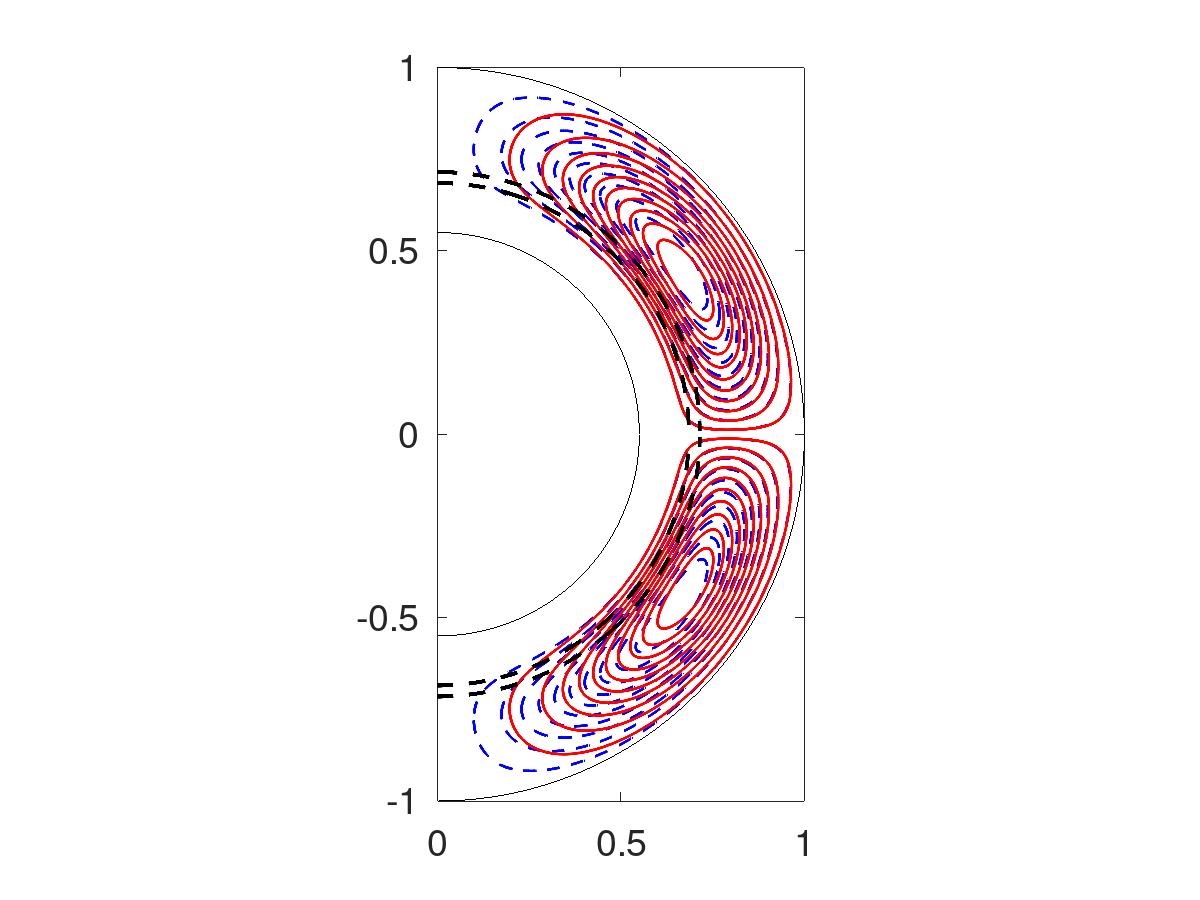}
\caption{The meridional circulation profiles used in the model. (Left) Shows the latitudinal variation of the surface \mc\ and 
(right) shows its streamlines in the 
CZ.
Red/solid and blue/dashed curves correspond to MC1 and MC2 profiles, respectively.
}
\label{fig:mc}
\end{figure*}

We take two profiles for the meridional flows which we call  MC1 and MC2. In MC1, we take $\epsilon = 3 $
and in MC2, we take $\epsilon = 2 $. These profiles are shown in \Fig{fig:mc}.
We note that MC1 is stronger in low latitudes and weaker in high latitudes. This MC1 profile has been used in the work presented in the Letter.  The profile MC2 has been used in our many previous publications \citep{KC11, KC16, KMB18}.

\begin{figure}
\includegraphics[scale=0.35]{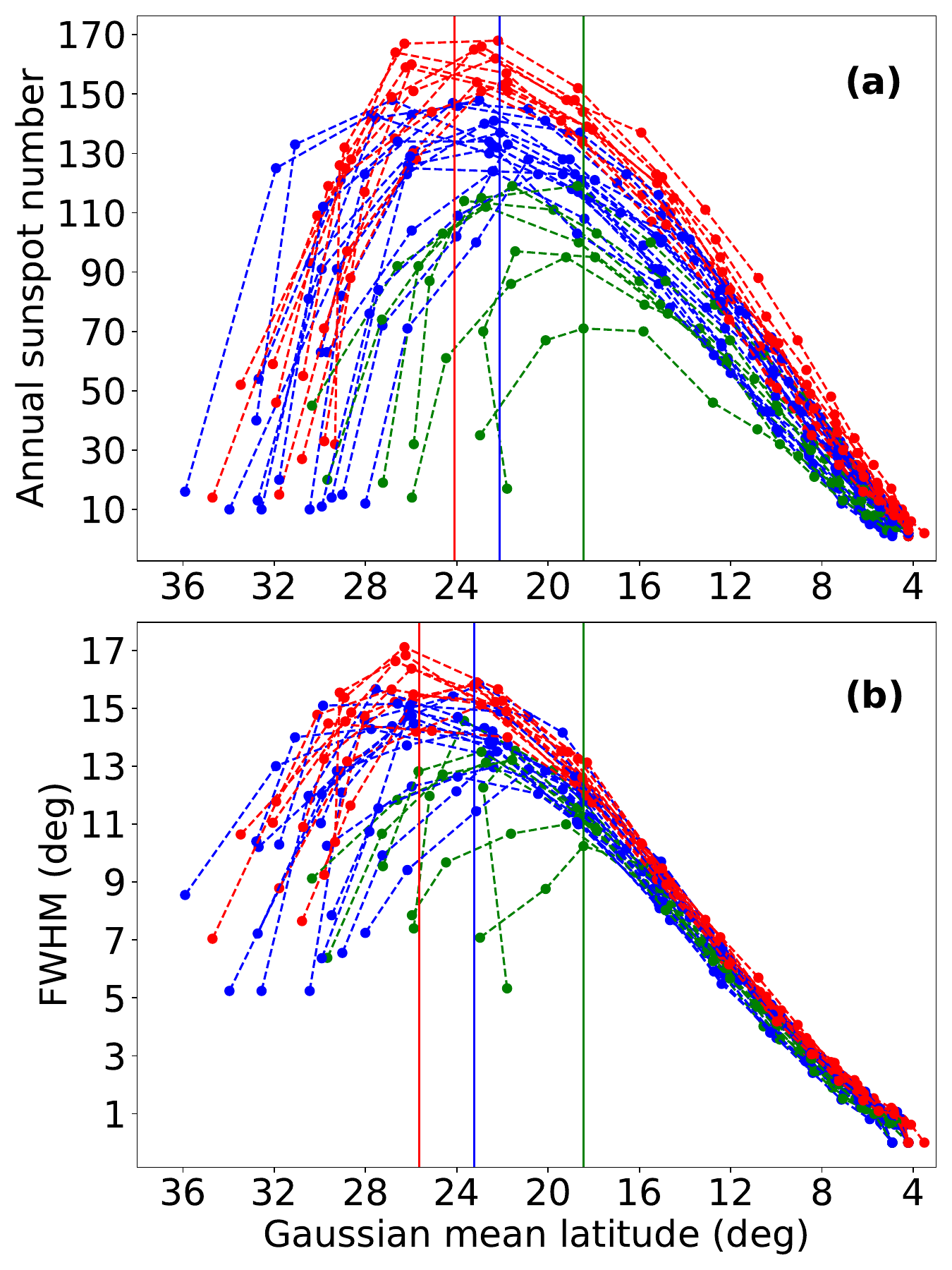}
\caption{
Results obtained from a dynamo run with MC2 profile for the meridional flow (as shown by the blue/dashed line in \Fig{fig:mc}).}
\label{fig:CSold}
\end{figure}

\vspace{-0.2 in}
\subsection{S2. Results with MC2 meridional flow}
\vspace{-0.15 in}
In \Fig{fig:CSold}, we show our results of the dynamo model in which the meridional circulation profile is changed to MC2 (see blue/dashed curves in \Fig{fig:mc} for its profile). In this case, we notice that the agreement with observation is slightly poor, particularly, the butterfly wing begin at somewhat higher latitudes and the activity level of not all cycles strictly decline at the same rates.

The reason for this poor agreement with observation is that the equatorward drift of the meridional flow at low latitudes is weak which is critical in shaping the trajectory of the sunspot band over the cycle.

\vspace{-0.2 in}
\subsection{S3. Computations of Cycle Parameters}
\vspace{-0.15 in}

We have explicitly studied the two aspects of \we, namely WE1 and WE2. WE1 refers to the anti-correlation between the rise time and the amplitude of the cycle, while WE2 refers to the positive correlation between the rise rate and the amplitude \citep{KC11}.  
To check these features, we need to compute the rise rate, rise time and amplitude of the cycle.
For this, we first smooth the three-month binned data using a Savitzky–Golay filter. Due to the overlap between the consecutive cycles, finding out the minimum is difficult. To avoid this problem we follow the same procedure as in \citet{KC11}. We take the rise time as the time taken by a cycle to grow its activity from $20\%$ to $80\%$ of its peak value. The rise rate is computed by dividing the difference between values of $80\%$ and $20\%$ strength by the rise time.

\vspace{-0.2 in}
\subsection{S4. Results with time delay in flux emergence}
\vspace{-0.15 in}

\begin{figure}
\includegraphics[scale=0.35]{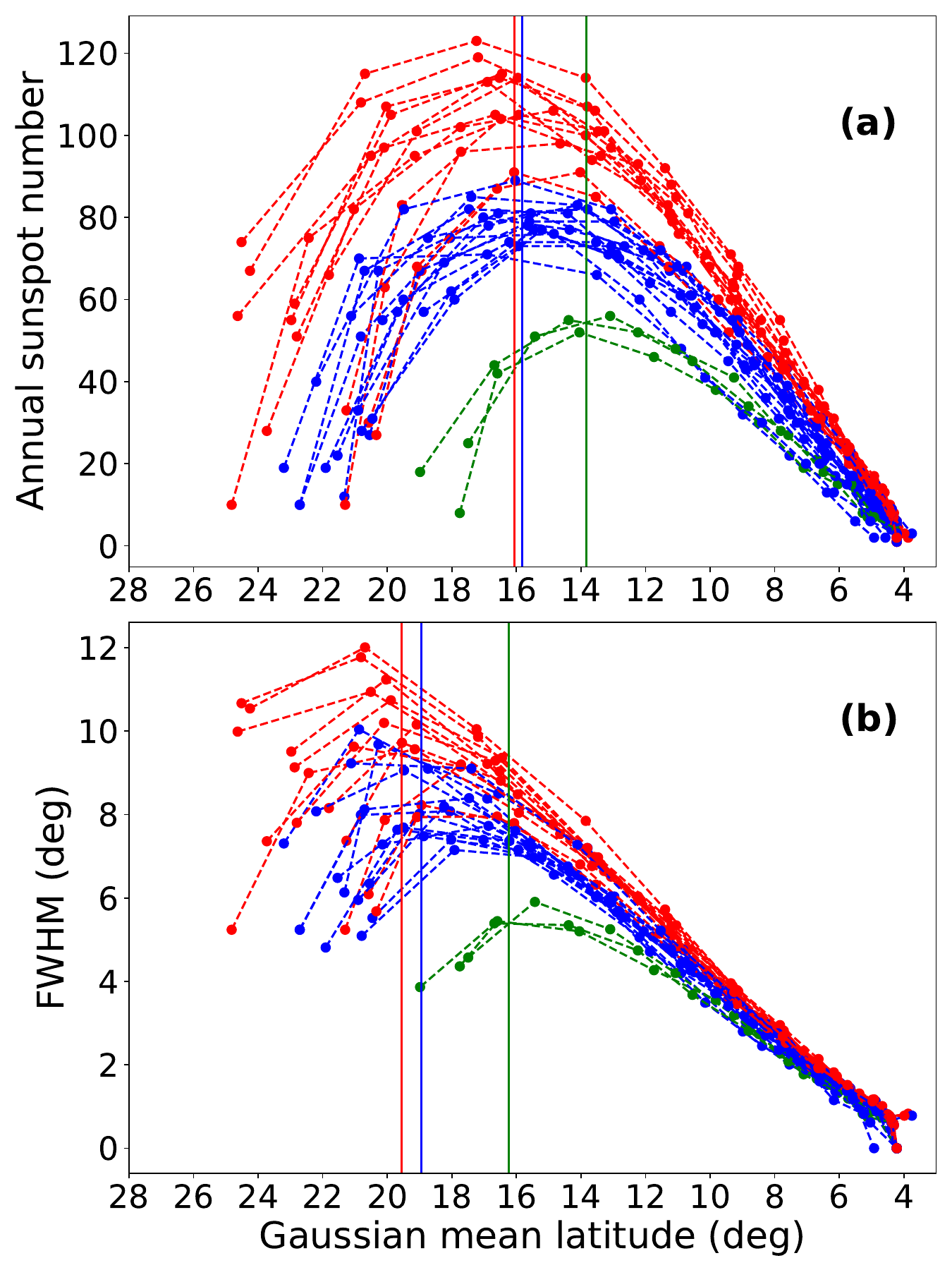}
\caption{Same as Figure~3 of the Letter but here a delay has been included in the flux eruption process.
}
\label{fig:delay}
\end{figure}

 So far in our model, we have ignored the delay induced due the rise of the magnetic flux from the BCZ to the surface.
 The magnetic buoyancy depends on the strength of the magnetic field at the BCZ and as a result the buoyancy induced delay in the source term of the poloidal field depends on magnetic field \citep{FFM94, Fournier18}. Here, we assume a simple inverse square relationship of the delay with the magnetic field strength such that the maximum delay is 6 months. The formula to incorporate this process in the code is following: $ D = 180 \times (B_c/B)^2$, where D is the number of days taken by the flux tube to rise from the BCZ to the surface (i.e. the delay in flux eruption), $B_c$ is the critical field strength to satisfy eruption condition and B is the value of toroidal field strength at the grid point where the eruption condition ($B> B_c$) has been met.
In our simulations 
we however do not observe a significant change in the magnetic cycle due to the magnetic field dependent delay (results shown in \Fig{fig:delay}). The reason for this is that in this model, once the magnetic field $B$ surpasses $B_c$ the flux loss $(B \rightarrow B/2$) condition does not allow the magnetic field to increase too much higher than $B_c$ and thus the buoyancy delay varies only a little.
However, it is interesting to note that on an average the mediocre and the weaker cycles (blue and green ones, respectively) shift  slightly towards  high latitude regions.

We note that the simulation presented here used the same initial condition that was used for the one presented in the Letter. The slight differences in the solutions (compare \Fig{fig:delay} with \Fig{fig:CS}) are due to the time delay introduced in the flux emergence.


We thank both the anonymous referees for providing insightful comments which helped us to improve the manuscript to a large extent.
A.B. and B.B.K. acknowledge financial support provided by ISRO/RESPOND (project No. ISRO/RES/2/430/19-20). RHC acknowledges the support of ERC Synergy Grant WHOLE SUN 810218.

\bibliographystyle{apsrev4-1}
\bibliography{paper}

\end{document}